# Quantum Noise of Kramers-Kronig Receiver

FAN ZHANG,[1, 4, 5] JIAYU ZHENG,[1] HAIJUN KANG,[2] FENGXIAO SUN,[3] QIONGYI HE,[3] AND XIAOLONG SU[2, 6]

[1] *State Key Laboratory of Advanced Optical Communication System and Networks, Frontiers Science Center for Nano-optoelectronics, School of Electronics, Peking University, Beijing 100871, China*

[2] *State Key Laboratory of Quantum Optics and Quantum Optics Devices, Institute of Opto-Electronics, Collaborative Innovation Center of Extreme Optics, Shanxi University, Taiyuan 030006, China*

[3] *State Key Laboratory for Mesoscopic Physics, School of Physics, Frontiers Science Center for Nano-optoelectronics & Collaborative Innovation Center of Quantum Matter, Peking University, Beijing 100871, China*

[4] *Peng Cheng Laboratory, Shenzhen 518055, China.*

[5] *fzhang@pku.edu.cn*

[6] *suxl@sxu.edu.cn*

**Abstract:** The Kramers-Kronig (KK) receiver provides an efficient method to reconstruct the complex-valued optical field by means of intensity detection given a minimum-phase signal. In this paper, we analytically show that for detecting coherent states through measuring the minimum-phase signal, while keeping the radial quantum fluctuation the same as the balanced heterodyne detection does, the KK receiver can indirectly recover the tangential component with fluctuation equivalently reduced to 1/3 times the radial one at the decision time, by adopting the KK relations to utilize the information of the physically measured radial component of other time of the symbol period. In consequence, the KK receiver achieves 3/2 times the signal-to-noise ratio of balanced heterodyne detection, while presenting an asymmetric quantum fluctuation distribution depending on the time-varying phase. Therefore, the KK receiver provides a feasible scheme to reduce the quantum fluctuation for obtaining the selected component to 2/ 3 times that of physically measuring the same component of the coherent state. This work provides a physical insight of the KK receiver and should enrich the knowledge of electromagnetic noise in quantum optical measurement.

## 1. INTRODUCTION

Optical receiver can be categorized as two basic schemes of coherent and direct detections [1]. Coherent receiver can either use homodyne or heterodyne detection to retrieve the complex full field information of the received signals. In contrast, direct detection can only obtain the signal intensity due to the square-law principle. By applying Kramers-Kronig (KK) relations [2, 3] to the field of optical communication, an advanced concept of KK receiver was proposed [4, 5], which can reconstruct the complex-valued optical field by means of intensity detection given a minimum-phase (MP) signal. Recently, the precision limits of quantum phase estimation through KK relations are analyzed in [6], where the joint estimation of two parameters following KK relations is investigated, indicating that the uncertainties of the calculated and the measured quantities are closely related. For the KK receiver applied on the MP signal, the amplitude waveform determines its corresponding phase waveform uniquely up to a constant phase offset. The physical mechanism is that the logarithm of the signal intensity (amplitude) and phase are related to each other through the KK relations apply to the signals that are causal in time. For a MP signal, it contains a data-carrying signal and a reference component that is a continuous-wave tone. When the spectrum of the signal's complex envelope is entirely above or below the reference frequency, which is so called single-sideband (SSB) signal, then a necessary and sufficient condition to be of minimum-phase is that its time-domain trajectory never encircles the origin of the complex plane. From the point of the configuration, the KK receiver is equivalent to heterodyne detection with one single photodetector (PD) [7]. By employing KK detection, high baud rate quadrature amplitude modulation (QAM) signal can be detected using a single PD at the receiver [7, 8].

Quantum noise, which comes from the uncertainty principle of quantum variables, sets a limitation for the signal-to-noise ratio ( *SNR* ) in the measurement of an optical signal. In quantum information with continuous variables, the information is encoded in the radial and tangential components of optical modes and measured by homodyne or heterodyne detection [9-15]. For heterodyne detection, a simultaneous measurement of two non-commuting quantum observables introduces excess noise that origins from vacuum fluctuations of the field, which imposes a fundamental limit on the *SNR* [16,17]. The KK receiver offers a joint measurement of the in-phase and quadrature components of the optical field, which corresponds to the radial and tangential components of an optical mode in quantum optics. A natural question is how about the quantum noise of the KK receiver?

In this paper, we study the quantum mechanical nature of the KK receiver and elucidate the fundamental relation between the KK receiver and the conventional balanced heterodyne detection. The result shows that the KK detection achieves 3/2 times the *SNR* of the balanced heterodyne detection. Interestingly, the quantum noise of the KK receiver shows an asymmetric fluctuation distribution, which is different from that of the balanced heterodyne detection. In this case, quantum noise of the KK receiver is time dependent for the in-phase and quadrature operators of an optical field, which corresponds to the radial and tangential components of an optical mode in quantum optics. Our result indicates the fluctuation properties of the quantum noise and its limit in the KK receiver when it is used to measure the coherent state.

## 2. QUANTUM MECHANICAL FUNDAMENTALS OF COHERENT STATE

For optical transmission, the information is conveyed by electromagnetic wave packets those are quantum states of the electromagnetic field. A signal carried by the coherent state $|\alpha_s\rangle$ is the eigen function of the photon annihilation operator $\hat{a}$, which is expressed as $\hat{a}|\alpha_s\rangle=\alpha_s|\alpha_s\rangle$. With the Hermitian conjugate operation, the creation operator satisfies $\langle\alpha_s|\hat{a}^\dagger=\alpha_s^*\langle\alpha_s|$, thus the photon number $n$ satisfies $\langle n_s\rangle=|\alpha_s|^2$. The non-Hermitian operator $\hat{a}$ can be separated into two Hermitian components $\hat{X}$ and $\hat{Y}$ by $\hat{a}=\hat{X}+j\hat{Y}$, which are the in-phase and quadrature field operators and satisfy the commutation relation given by $[\hat{X},\hat{Y}]=j/2$. Here $\hat{X}=\sqrt{\frac{\omega}{2\hbar}}\hat{q}$ and $\hat{Y}=\frac{\hat{p}}{\sqrt{2\hbar\omega}}$ are essentially dimensionless position and momentum operators, which correspond to the radial and tangential components of an optical mode in quantum optics, respectively.

The Heisenberg uncertainty principle sets an upper limit on the precision of a quantum measurement. We assume that $\langle(\Delta\hat{M})^2\rangle$ represents the quantum variance of measuring the observable $\hat{M}$. The coherent state is one of the minimum uncertainty states, of which the quantum variances of the in-phase and quadrature operators are $\langle(\Delta\hat{X})^2\rangle=1/4$ and $\langle(\Delta\hat{Y})^2\rangle=1/4$, respectively. Thus, the total variance is $\langle(\Delta\hat{a})^2\rangle=\langle(\Delta\hat{X})^2\rangle+\langle(\Delta\hat{Y})^2\rangle=1/2$ [16-18].

The quantum variances of the radial and tangential of a coherent state are $\langle(\Delta|\hat{a}|)^2\rangle=1/4$ and $\langle(\Delta\hat{\phi})^2\rangle=1/(4\langle n\rangle)$, respectively [18]. Here the photon number $n$ satisfies $\langle n\rangle=|\alpha|^2$. For any coherent state $|\alpha\rangle$ that has a time dependence $e^{-j\omega t}$, its corresponding annihilation operator is defined as $\hat{A}|\alpha\rangle=e^{-j\omega t}\alpha|\alpha\rangle$. $\hat{A}$ and its conjugate operator $\hat{A}^\dagger$ obey the commutation relation $[\hat{A},\hat{A}^\dagger]=1$.

## 3. QUANTUM MECHANICS MODEL OF BALANCED HETERODYNE DETECTION

As shown in Fig. 1 (a), the balanced heterodyne detection is adopted to get the linear beating between the local oscillator (LO) $\hat{A}_L$ at frequency $\omega_L$ and the signal $\hat{A}_s$ at frequency $\omega_s$. The waves $\hat{B}_1$ and $\hat{B}_2$ that incident upon the two PDs are

$$\hat{B}_1=\frac{1}{\sqrt{2}}[\hat{A}_L+\hat{A}_s'+\hat{A}_i'-j(\hat{A}_s+\hat{A}_i)], \tag{1}$$

$$\hat{B}_2=\frac{1}{\sqrt{2}}[-j(\hat{A}_L+\hat{A}_s'+\hat{A}_i')+(\hat{A}_s+\hat{A}_i)]. \tag{2}$$

Here $\hat{B}_1$ and $\hat{B}_2$ have considered the signal $\hat{A}_s$ at frequency $\omega_s$ and its image $\hat{A}_i$ at $\omega_i=2\omega_L-\omega_s$, as well as the vacuum fluctuations $\hat{A}_s'$ and $\hat{A}_i'$ at these two frequency points brought by LO [19-24] as shown in Fig. 1 (a). Note that $n$ photons are received by the photodetector within the signal duration $T$. With the elementary charge $q$, the current is $\hat{I}=q\hat{n}/T$. The difference between the current collected by the two detectors [25-26] is

$$\begin{aligned}\langle\hat{I}(t)\rangle_{Bal}&=k\langle\hat{B}_2^\dagger\hat{B}_2-\hat{B}_1^\dagger\hat{B}_1\rangle\\&=2k\left|\alpha_s\alpha_L^*\right|\sin(\omega_{IF}t-\arg(\alpha_s\alpha_L^*)).\end{aligned} \tag{3}$$

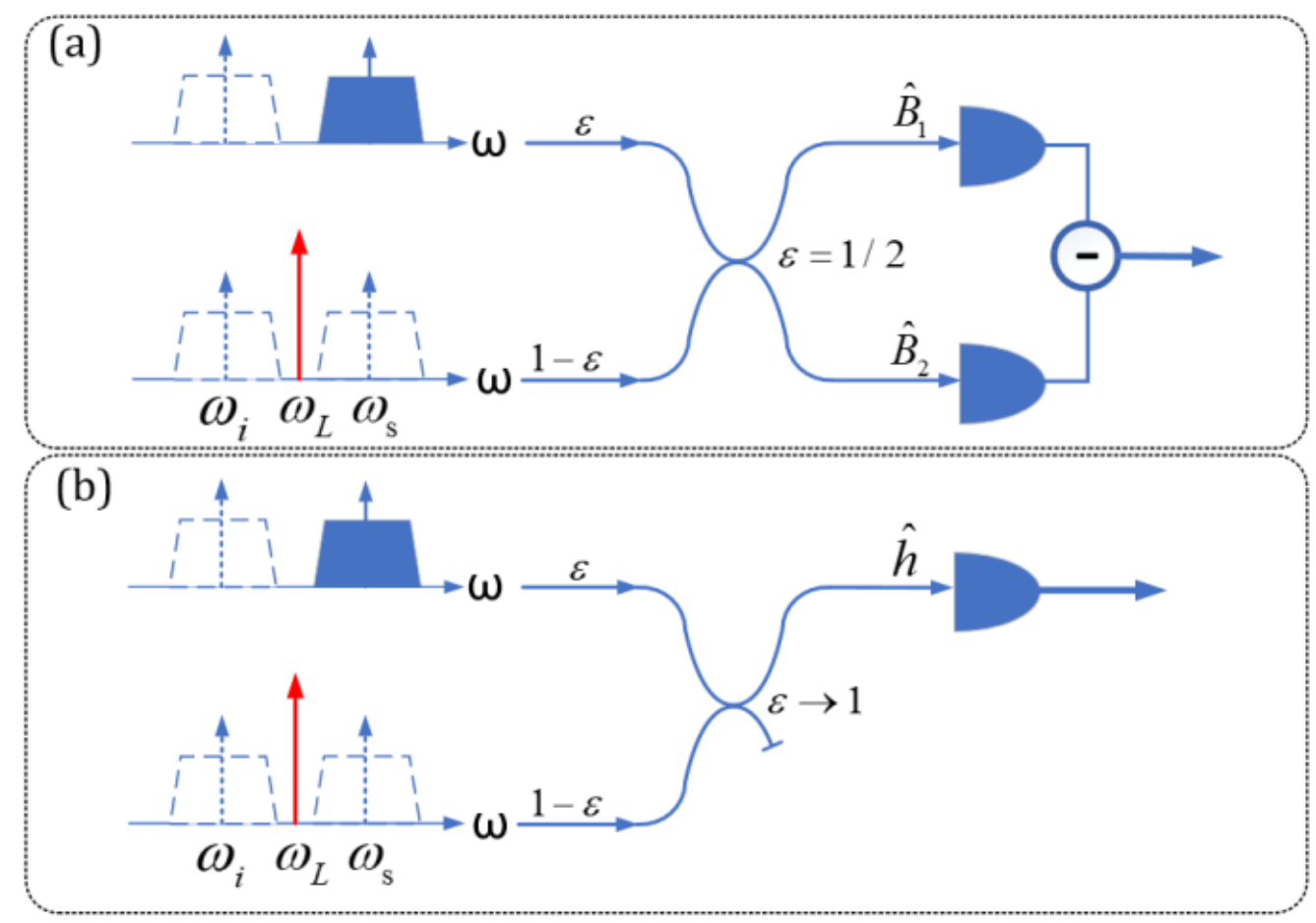

**Fig. 1.** Basic configurations of the balanced heterodyne and KK detection. (a) Balanced heterodyne detection; (b) The KK detection.

Here $k=q/T$. $\alpha_s$ and $\alpha_L$ are the eigenvalues for the signal and the LO, respectively. $\omega_{IF}=\omega_s-\omega_L$ is the intermediate frequency. For the image band, there exits that $\langle\hat{A}_i^\dagger\hat{A}_i\rangle=0$. Yet the presence of $\hat{A}_i$ contributes to the fluctuations. The variance of photocurrent is given by

$$\langle(\Delta\hat{I})^2\rangle_{Bal}=\langle\hat{I}(t)^2\rangle-\langle\hat{I}(t)\rangle^2=k^2(3\langle n_s\rangle+2\langle n_L\rangle). \tag{4}$$

Then the $S/N$ of the balanced heterodyne detection is calculated from Eq. (3) and (4), which is given by

$$(S/N)_{Bal}=\frac{\overline{(\langle\hat{I}(t)\rangle_{Bal})^2}}{\langle(\Delta\hat{I})^2\rangle_{Bal}}=\frac{2\langle n_s\rangle\langle n_L\rangle}{2\langle n_L\rangle+3\langle n_s\rangle}=\frac{\langle n_s\rangle}{1+3\langle n_s\rangle/2\langle n_L\rangle}. \tag{5}$$

Here the operation "$\overline{\hat{M}(t)}$" means averaging the value $\hat{M}(t)$ by time.

## 4. QUANTUM MECHANICS MODEL OF THE KK RECEIVER

For the KK receiver, the signal is retrieved with one single PD of the balanced detection, which is equivalent to sending the LO along with the signal from the transmitter side as shown in Fig. 1 (b). The MP signal is constructed with a co-polarized reference carrier with real-valued amplitude and a complex optical data. When the MP signal impinges upon one single PD, the beating between the reference carrier and the data-carrying signal is analogue to heterodyne detection.

Without loss of generality, we express the MP signal as $h(t)=\sqrt{1-\varepsilon}A_L+\sqrt{\varepsilon}A_s(t)$ depicted in Fig. 1 (b), where the power transmission $\varepsilon$ is set to $\frac{1}{2}$ for balanced detection in Fig.1 (a). For the KK receiver, to maximize the utilization of signals, we set the power transmission $\varepsilon \to 1$. Considering the fast time-vary phase of the optical carrier, we rewrite the MP signal as $h(t)=[\sqrt{1-\varepsilon}\alpha_L+\sqrt{\varepsilon}\alpha_s(t)e^{-j\omega_{IF}t}]e^{-j\omega_L t}$. Here only the expectation values are considered as in the classical KK receiver proposals [4, 5], where the effects of the quantum fluctuations will be included latter. For a complex data-carrying signal $\alpha_s(t)$ whose spectrum is contained in the frequency range $(0,W]$, an ideal SSB signal that has its spectrum closely above the reference frequency $\omega_L$ satisfies $\omega_{IF}=\pi W$. The second-order term $\varepsilon|\alpha_s(t)|^2$ is commonly referred to as the signal-to-signal beat interference (SSBI) [4], whose spectrum is in the frequency of $[-W/2,W/2]$. The balanced heterodyne detection can eliminate SSBI by the subtraction in Eq. (3). In contrast, for the KK receiver that with one single PD, the SSBI cancellation is not required as $\varepsilon|\alpha_s(t)|^2$ is an essential part of the detected current intensity that required for the reconstruction [4].

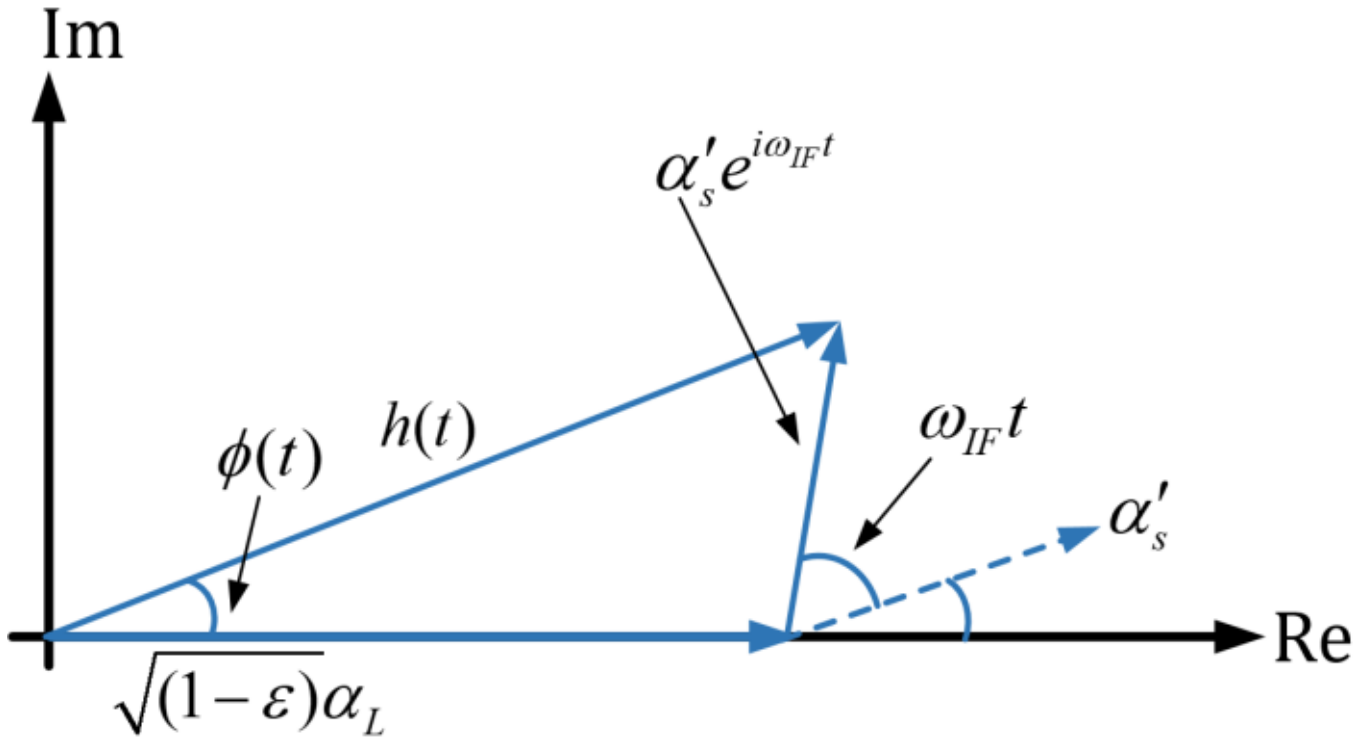

**Fig. 2.** The relationship between the retrieved $\alpha_s'$ and $h(t)$ for KK detection.

For each signal transmitted, the purpose of KK receiver is to recover the baseband signal $\alpha_s(t)$ at the decision time $t$ from the amplitude and phase of $h(t)$. The current $I(t)=k|h(t)|^2$ is obtained from a single PD. With the MP condition $\varepsilon\langle n_s\rangle<(1-\varepsilon)\langle n_L\rangle$, the phase of $h(t)$ that can be extracted from the current logarithm with Hilbert transform is defined as [5]

$$\phi(t)=\frac{1}{2\pi}\mathcal{P}\int_{-\infty}^{+\infty}\frac{\ln|I(t')|}{t-t'}dt'. \tag{6}$$

Here $\mathcal{P}$ represents the Cauchy principal value. The complex signal field $\alpha_s$ is retrieved as $\alpha_s'$.

$$\alpha_s'=\left[\frac{1}{\sqrt{\varepsilon}}|h(t)|e^{j\phi(t)}-\frac{\sqrt{1-\varepsilon}}{\sqrt{\varepsilon}}\sqrt{\langle n_L\rangle}\right]e^{j\omega_{IF}t}. \tag{7}$$

The relation between $\alpha_s'$ and $h(t)$ is shown in Fig. 2, in which $\phi(t)$ is the time-varying phase difference between $h(t)$ and $\sqrt{1-\varepsilon}\alpha_L$.

From Eq. (7), the relation between $\Delta\hat{\alpha}_s'$ and $\Delta\hat{h}$ is given by multiplying the phase factor $e^{j\omega_{IF}t}$.

$$\Delta\hat{\alpha}_s'=\Delta\hat{h}e^{j\omega_{IF}t}. \tag{8}$$

It is obvious that the retrieved $\alpha_s'$, which can be obtained from $h(t)$ after a phase shift with an exponent translation, has the quantum noise determined by that of $h(t)$. Therefore, we study the quantum fluctuations of $\alpha_s'$ through those of $h(t)$ in the case of KK detection.

Considering the signal and its image, as well as their corresponding vacuum fluctuations brought by the carrier [19-24], we have $\hat{h}=\sqrt{1-\varepsilon}(\hat{A}_L+\hat{A}_s'+\hat{A}_i')+\sqrt{\varepsilon}(\hat{A}_s+\hat{A}_i)$ before the PD port as shown in Fig. 1 (b), in which the operators $\hat{A}_L$, $\hat{A}_s'$, $\hat{A}_i'$, $\hat{A}_s$ and $\hat{A}_i$ are all for single frequency.

Note that $I(t)$ corresponds to the photocurrent expectation $\langle\hat{I}(t)\rangle_{KK}$, which is obtained by projection via the coherent states, product states of the LO and the signal (and its image) states.

$$\begin{aligned}\langle\hat{I}(t)\rangle_{KK}&=k\langle\hat{h}^\dagger\hat{h}\rangle=k\langle\alpha_L|\langle\alpha_s|\langle\alpha_i|\hat{h}^\dagger\hat{h}|\alpha_i\rangle|\alpha_s\rangle|\alpha_L\rangle\\&=k[\varepsilon\langle n_s\rangle+(1-\varepsilon)\langle n_L\rangle\\&\quad+2\sqrt{\varepsilon(1-\varepsilon)}|\alpha_s\alpha_L^*|\cos(\omega_{IF}t-\arg(\alpha_s\alpha_L^*))]\end{aligned} \tag{9}$$

The mean variance of the photocurrent is given by

$$\begin{aligned}\langle(\Delta\hat{I}(t))^2\rangle_{KK}&=\langle\hat{I}(t)^2\rangle-\langle\hat{I}(t)\rangle^2\\&=2k^2[\varepsilon\langle n_s\rangle+(1-\varepsilon)\langle n_L\rangle\\&\quad+2\sqrt{(1-\varepsilon)\varepsilon}|\alpha_s\alpha_L^*|\cos(\omega_{IF}t-\arg(\alpha_s\alpha_L^*))]\end{aligned}. \tag{10}$$

The fluctuations of the photon current are related to the operators $\hat{A}_i\hat{A}_i^\dagger$, $\hat{A}_s\hat{A}_s^\dagger$ and $\hat{A}_L\hat{A}_L^\dagger$, which are in reverse order to the photon number operator. It is worth noting that the ratio of $\langle(\Delta\hat{I}(t))^2\rangle_{KK}$ to $\langle\hat{I}(t)\rangle_{KK}$ is a constant $2k$. A general mathematical proof is given in the Supplement, Note 1.

The operator $\Delta\hat{\phi}(t)$ representing the difference between the measured phase $\hat{\phi}(t)$ and the expected phase $\phi(t)$, which is obtained after Taylor expansion and approximation of the logarithm function in Eq. (6), is expressed by

$$\Delta\hat{\phi}(t)=\hat{\phi}(t)-\phi(t)\approx\frac{1}{2\pi}\mathcal{P}\int_{-\infty}^{+\infty}\frac{1}{t-t'}\left[\frac{\hat{I}(t')-I(t')}{I(t')}\right]dt'. \tag{11}$$

The phase fluctuations $\langle(\Delta\hat{\phi}(t))^2\rangle$ can be expressed in discrete form with $t=l\delta t$, $t'=m\delta t$ and $dt'\to\delta t$. For each $m$, $\langle(\Delta\hat{I}(t'))^2\rangle/I(t')^2$ with $t'\in[(m-1)\delta t,m\delta t]$ can be represented by $\langle(\Delta\hat{I}(m\delta t))^2\rangle/I(m\delta t)^2$.

$$\begin{aligned}\langle(\Delta\hat{\phi}(t))^2\rangle&=\lim_{\delta t\to 0}\frac{1}{4\pi^2}\left[\sum_{m=-\infty,m\neq l}^{m=+\infty}\frac{1}{(l-m)^2}\frac{\langle(\Delta\hat{I}(m\delta t))^2\rangle}{I(m\delta t)^2}\right]\\&=\lim_{\delta t\to 0}\frac{1}{12}\frac{\langle(\Delta\hat{I}(t))^2\rangle}{(I(t))^2}=\frac{k}{6I(t)}\end{aligned} \tag{12}$$

In the derivation of Eq. (12), the cross terms vanish due to the fact that $\Delta\hat{I}(m_1\delta t)$ and $\Delta\hat{I}(m_2\delta t)$ at different instants are irrelevant.

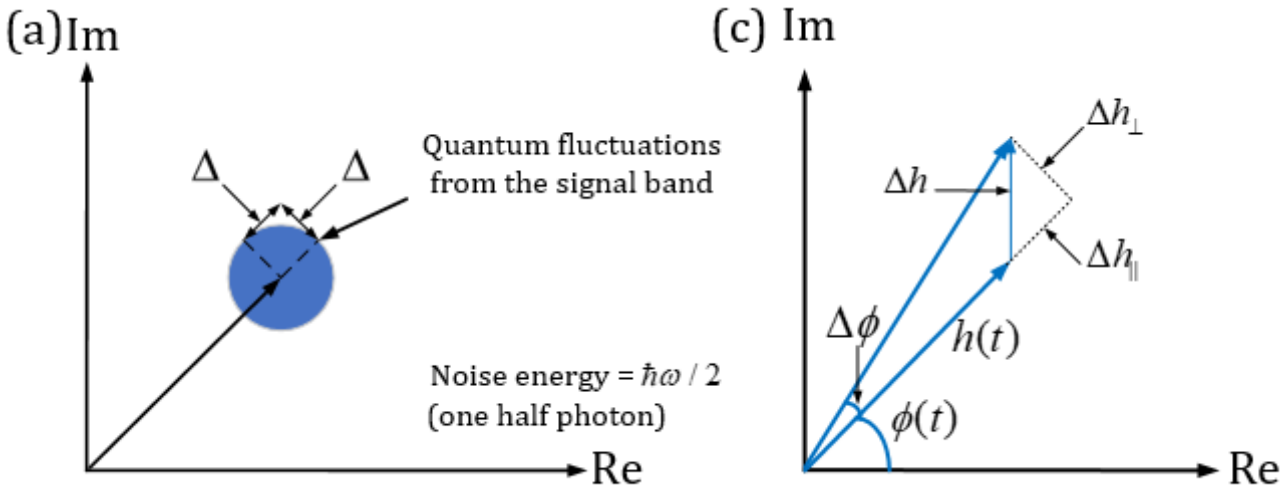

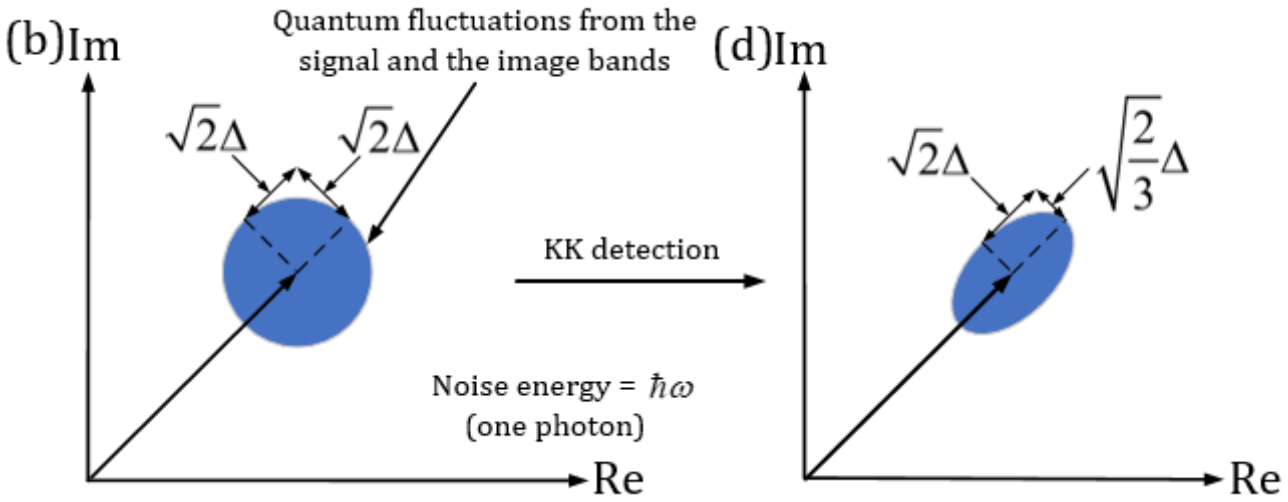

**Fig. 3.** Quantum noise associated with the KK receiver. (a) Quantum fluctuations from the signal band for homodyne detection; (b) Quantum fluctuations from the signal and the image bands for heterodyne detection; (c) The radial and the tangential fluctuations of the MP signal; (d) Phasor diagram of the light after KK detection. $\Delta = 1/2$ represents the standard deviation of the quantum fluctuations of the coherent state.

Therefore, only the square terms remain. To get the result of Eq. (12), Eq. (9) and Eq. (10) are applied, and the fact that $\sum_{m=1}^{+\infty} 1/m^2 = \pi^2/6$ and $1/I(t)$ is continuous and derivative are considered. A detailed derivation of Eq. (12) is given in the Supplement, Note 2. Note that a similar analysis of the quantum fluctuations of the quantities with KK relation can be found in [6].

Assuming "$\parallel$" and "$\perp$" respectively represent the radial and tangential components corresponding to the amplitude and phase of $h(t)$, the $\langle(\Delta\hat{h}_{\parallel})^2\rangle$ and $\langle(\Delta\hat{h}_{\perp})^2\rangle$ subsequently represent the radial and the tangential fluctuations of $h(t)$, which correspondingly have the form $\Delta\hat{h}_{\parallel} = \Delta\left|\hat{h}(t)\right|$ and $\Delta\hat{h}_{\perp} = \left|h(t)\right|\Delta\hat{\phi}(t)$ as shown in Fig. 3 (c). We thus obtain the following quantum mechanical variances for the detection of a coherent state.

$$\langle(\Delta\hat{h}_{\parallel})^2\rangle = \langle(\Delta\left|\hat{h}(t)\right|)^2\rangle = \frac{\langle[\sqrt{\hat{I}(t)}-\sqrt{I(t)}]^2\rangle}{k} \approx \frac{1}{k}\langle[\frac{\hat{I}(t)-I(t)}{2\sqrt{I(t)}}]^2\rangle = \frac{\langle(\Delta\hat{I}(t))^2\rangle}{4kI(t)} = \frac{1}{2}, \qquad (13)$$

$$\langle(\Delta\hat{h}_{\perp})^2\rangle = \frac{1}{k}\langle I(t)(\Delta\hat{\phi}(t))^2\rangle = \frac{\langle(\Delta\hat{I}(t))^2\rangle}{12kI(t)} = \frac{1}{6}. \qquad (14)$$

From Eq. (8), the quantum fluctuation $N$ of $\alpha_s'$ is calculated as

$$N = \langle(\Delta\hat{\alpha}_s')^2\rangle = \langle(\Delta\hat{h})^2\rangle\,|e^{j\omega_{IF}t}|^2 = \langle(\Delta\hat{h}_{\parallel})^2\rangle + \langle(\Delta\hat{h}_{\perp})^2\rangle = \frac{2}{3}. \qquad (15)$$

Note that for a coherent state, the total quantum fluctuation $N_C = \langle(\Delta\hat{X})^2\rangle + \langle(\Delta\hat{Y})^2\rangle = 1/2 < 2/3$, which indicates that the Heisenberg uncertainty principle is not violated in our KK detection scheme.

Considering Eq. (15), the upper limit of $S/N$ of KK detection due to the quantum fluctuation is equal to $\frac{3}{2}\langle n_s\rangle$.

$$\left(S/N\right)_{KK} = \langle n_s\rangle / N = \frac{3}{2}\langle n_s\rangle. \qquad (16)$$

In contrast, for the balanced heterodyne detection, the $\left(S/N\right)_{Bal}$ approaches $\langle n_s\rangle$ [25-26] when $\langle n_L\rangle \gg \langle n_s\rangle$ as shown in Eq. (5).

From Eq. (13) and Eq. (14) we find that, with the KK relations, $\hat{h}_{\parallel}$ has the radial fluctuations twice that of a coherent state, as indicated in Fig. 3(a) and Fig. 3(d), which embodies the contribution of the signal $\hat{A}_s$ as well as the image $\hat{A}_i$. Meanwhile, the tangential fluctuations resulted by KK detection are only 1/3 times of the radial ones. This ratio reveals the physical contribution of KK relation, which is, physically, from the Hilbert transform. The introduction of the Hilbert transform provides extra information for measuring the tangential component of the MP signal compared with direct measurement. It is the smaller tangential fluctuation that leads to a greater $SNR$ of the KK detection than that of the balanced heterodyne detection.

The Arthurs-Kelly model [27-29] points out that, in order to realize the simultaneous measurement of a pair of non-reciprocal mechanical observables, two new reciprocal observables can be constructed, which have the same expected values as the former two do. Meanwhile, the total fluctuations resulted by the measurement process of the two observables constructed will be at least doubled compared with the respective measurement results of the two original observables. It can be proved that when $\langle n_L\rangle \gg \langle n_s\rangle$ holds, for the balanced heterodyne detection, as a pair of reciprocal observables constructed, the in-phase and quadrature components of $\hat{A}_s + \hat{A}_i^{\dagger}$ can be measured simultaneously, which contribute the same to the total fluctuations.

Different from the balanced heterodyne detection, for each signal transmitted, the KK receiver makes use of the detected intensity information of photocurrent in the signal duration, to measure (calculate) the MP signal phase $\phi(t)$. Instead of measuring the tangential component directly, KK receiver utilizes the Kramers-Kronig constraint relations between the expected value of phase at the decision time $\phi(t)$, and that of photocurrent intensity at other times $I(t')$ ( $-\infty < t' < +\infty$ , $t' \neq t$ ). This introduces more knowledge about phase $\phi(t)$ provided by the Hilbert transform, into the detection process, further reducing the fluctuations of measuring $\hat{h}_{\perp}$ to only 1/3 times those of measuring $\hat{h}_{\parallel}$. Therefore, through the KK receiver, the amplitude and phase of MP signal can be obtained at the same time point, while the fluctuation of measuring tangential component can be reduced to 1/6, which is even lower than 1/4 of directly measuring the same component of a coherent state signal. This further leads to that the total quantum variance of KK receiver is only 2/3 compared with 1 of balanced heterodyne detection. Then using KK relations, the $SNR$ of the coherent state signals can achieve $\frac{3}{2}\langle n_s\rangle$, which is higher than $\langle n_s\rangle$ of balanced heterodyne detection. It should be noted that, in KK detection, only the intensity is physically measured, while the phase is calculated from the intensity through the KK relations. Therefore, the reduced tangential quantum fluctuation 1/6 (less than 1/4 of a coherent state signal) is obtained through the mathematical calculation instead of a realistic physical measurement. The conclusion does not violate Heisenberg uncertainty principle.

As shown in Fig. 3 (b) and Fig. 3 (d), the KK receiver results in a smaller total noise energy than the conventional balanced heterodyne detection does, while the end point of the MP signal from the KK receiver has an elliptical asymmetrical distribution, indicating a specific physical essence of KK detection in quantum case.

Now we check the quantum variances of the in-phase and quadrature field operators of the signal $\langle(\Delta\hat{\alpha}'_{s1})^2\rangle$ and $\langle(\Delta\hat{\alpha}'_{s2})^2\rangle$, which represent the variances of position and momentum components of $\hat{\alpha}'_s$ respectively. From Fig. 2, by expanding $\Delta\hat{\alpha}'_s$ and $\Delta\hat{h}$, we can rewrite Eq. (8) as

$$\begin{aligned}\Delta\hat{\alpha}'_{s1}+j\Delta\hat{\alpha}'_{s2}=&(\Delta\hat{h}_{\parallel}\cos(\phi(t))-\Delta\hat{h}_{\perp}\sin(\phi(t)))e^{j\omega_{IF}t}\\&+j(\Delta\hat{h}_{\parallel}\sin(\phi(t))+\Delta\hat{h}_{\perp}\cos(\phi(t)))e^{j\omega_{IF}t}\end{aligned}. \quad (17)$$

Substituting the relation $e^{j\omega_{IF}t}=\cos\omega_{IF}t+j\sin\omega_{IF}t$ into Eq. (17) to extract the real and the imaginary parts, respectively, we get

$$\begin{bmatrix}\Delta\hat{\alpha}'_{s1}\\ \Delta\hat{\alpha}'_{s2}\end{bmatrix}=\begin{bmatrix}\cos(\omega_{IF}t+\phi(t)) & -\sin(\omega_{IF}t+\phi(t))\\ \sin(\omega_{IF}t+\phi(t)) & \cos(\omega_{IF}t+\phi(t))\end{bmatrix}\begin{bmatrix}\Delta\hat{h}_{\parallel}\\ \Delta\hat{h}_{\perp}\end{bmatrix} \quad (18)$$

Then the variances of $\langle(\Delta\hat{\alpha}'_{s1})^2\rangle$ and $\langle(\Delta\hat{\alpha}'_{s2})^2\rangle$ are calculated as

$$\begin{bmatrix}\langle(\Delta\hat{\alpha}'_{s1})^2\rangle\\ \langle(\Delta\hat{\alpha}'_{s2})^2\rangle\end{bmatrix}=\begin{bmatrix}\cos^2(\omega_{IF}t+\phi(t)) & \sin^2(\omega_{IF}t+\phi(t))\\ \sin^2(\omega_{IF}t+\phi(t)) & \cos^2(\omega_{IF}t+\phi(t))\end{bmatrix}\begin{bmatrix}\langle(\Delta\hat{h}_{\parallel})^2\rangle\\ \langle(\Delta\hat{h}_{\perp})^2\rangle\end{bmatrix} \quad (19)$$

Eq. (19) is obtained using the relation $\langle\Delta\hat{h}_{\parallel}\Delta\hat{h}_{\perp}\rangle=0$. Note that '$T$' stands for matrix transposition. Eq. (19) shows that $(\langle(\Delta\hat{\alpha}'_{s1})^2\rangle,\langle(\Delta\hat{\alpha}'_{s2})^2\rangle)^T$ are related to $(\langle(\Delta\hat{h}_{\parallel})^2\rangle,\langle(\Delta\hat{h}_{\perp})^2\rangle)^T$ through a transformation matrix determined by $\omega_{IF}t+\phi(t)$, which brings no physical discrepancy between the fluctuation distributions of $\hat{\alpha}'_s$ and $\hat{h}$. As shown in Eq. (13) and Eq. (14), three times difference between $\langle(\Delta\hat{h}_{\parallel})^2\rangle$ and $\langle(\Delta\hat{h}_{\perp})^2\rangle$ results in the asymmetric distribution of $\langle(\Delta\hat{\alpha}'_s)^2\rangle$. For the retrieved $\hat{\alpha}'_s$, with the total fluctuations kept the same as those of $\hat{h}$, the in-phase and quadrature fluctuations both have a time dependent evolution as a function of $\omega_{IF}t+\phi(t)$. In other words, there exists a time varying accuracy for the measured position and momentum components of a coherent signal using the KK receiver. From Fig. 2, we can see that for a given $\alpha_s$, the time varying of the projected in-phase and quadrature field fluctuation is related to the phase evolution of $\omega_{IF}t$. The time-varying property of $(\langle(\Delta\hat{\alpha}'_{s1})^2\rangle,\langle(\Delta\hat{\alpha}'_{s2})^2\rangle)$ essentially results from the factor $e^{j\omega_{IF}t}$ in Eq. (7), in which $t$ as the decision time is selected artificially. This means that based on measuring $\hat{h}$ through the KK receiver, additional mathematical constraints are added to further recover $\alpha'_s$.

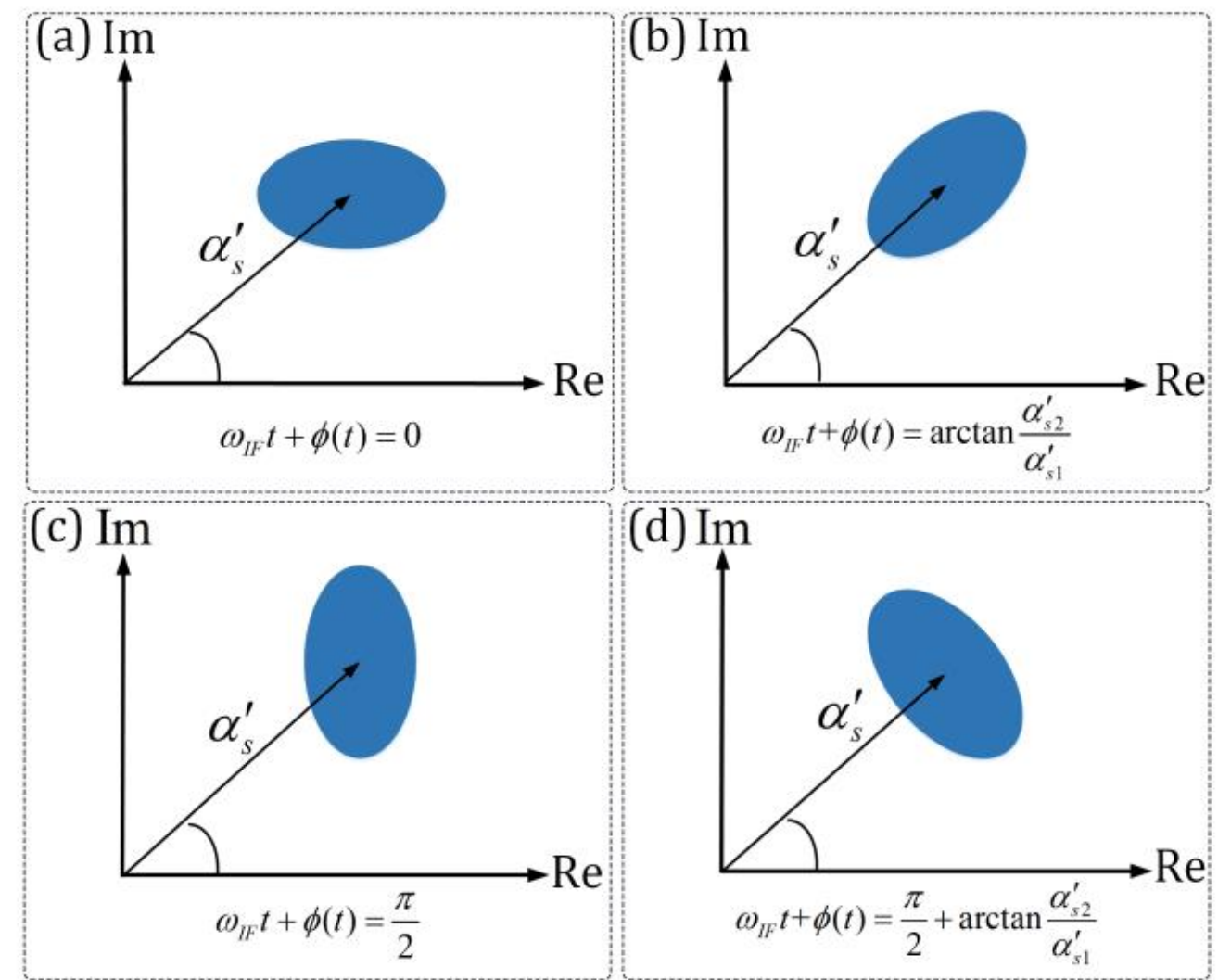

**Fig. 4.** Phasor diagram of the retrieved signal $\alpha'_s$ at four specific orientations for a coherent state. The quantum noise of the coherent state shows an asymmetric fluctuation distribution in the KK receiver, which can be reduced to lower than the vacuum fluctuation in a selected direction (orthogonal to the directly measured one).

For instance, Fig. 4 shows the phasor diagram of the retrieved signal $\alpha'_s$ at four specific orientations for a coherent state. In contrast to the conventional balanced heterodyne or homodyne detection, the KK receiver exhibits asymmetric quantum noise in both in-phase and quadrature operators that depends on the time varying phase of $\omega_{IF}t+\phi(t)$ when it is applied to measure a coherent state.

Combing Fig. 2 and Fig. 4, it can be seen that when $\sqrt{1-\varepsilon}\alpha_L\gg\alpha_s$ holds, there will hold $\phi(t)\approx 0$ no matter how $t$ varies. Consequently, the phase term $\omega_{IF}t+\phi(t)$ will be approximately equal to $\omega_{IF}t$. Since $t$ is defined as the decision time while $\omega_{IF}$ represents the intermediate frequency, both of which are known varieties. In principle, we can control the fluctuation distribution by reasonably selecting these two variables. Therefore, the quantum fluctuation of the electromagnetic field can be reduced to 1/6 in the specific direction selected to be measured more accurately.

## 5. SIMULATION RESULTS

To validate our analytical derivation and illustrate the statistical nature of quantum fluctuations of KK detection, we perform numerical simulation with up to 40000 signals which are detected with the KK receiver. Without losing generality, the Quadrature Phase Shift Keying (QPSK) modulation format is selected in the simulation for simplicity.

To approximate the KK relations, Eq. (6) is discretized into the form of Riemann sum. Considering that $\omega_s$ and $\omega_L$ are both for single frequency, $\omega_{IF} = \omega_s - \omega_L$ can be set small enough to meet the condition $\delta t \to 0$ (relative to $T_{IF}$ defined as $T_{IF} = 2\pi / \omega_{IF}$ ) in Eq. (12). This is ensured by selecting $2\times10^3$ sampling points for $T_{IF}$ to realize $\delta t = T_{IF} / (2\times10^3)$. Meanwhile, up to $2\times10^5$ sampling points are selected for $T$ as the integral interval in Eq. (6). While the decision time $t$ is set as $t = T/2$ to keep the integral intervals on both sides of $t$ symmetrical.

Since the Poisson distribution can be approximated by Gaussian distribution as its mean becomes large [30], Gaussian random number is used to simulate the measured noise. At the same time, by adding the current noise at each time point independently, the Gaussian white noise property is satisfied. Thus, to model the fluctuations of current $\langle(\Delta\hat{I}(t))^2\rangle_{KK}$, as has been proved in the Supplement, Eq. (S5), we may as well assume that the current deviation $\Delta I(t)_{KK}$ has the following form for carrying out the simulation using MATLAB.

$$\Delta I(t)_{KK} = \sqrt{2k\langle\hat{I}(t)\rangle_{KK}} \times randn\,. \tag{20}$$

Where '*randn*' represents the Gaussian random number with the mean value as 0 and the standard deviation as 1, thus the quantum noise of $I(t)_{KK}$ at time point $t$ is modeled as the independent Gaussian noise with $0$ mean and variance equaling to $2k\langle\hat{I}(t)\rangle_{KK}$.

The carrier to signal power ratio ( $CSPR$ ) of the MP signal can be defined as

$$CSPR = \frac{|\sqrt{1-\varepsilon}\alpha_L|^2}{|\sqrt{\varepsilon}\alpha_s|^2} = \frac{(1-\varepsilon)\langle n_L\rangle}{\varepsilon\langle n_s\rangle} > 1\ . \tag{21}$$

First, we set $CSPR$ to 10 dB and let $\langle n_s\rangle$ start from 20 and increase in a step of 20. For QPSK signals, it can be seen that the fluctuations surrounding each constellation point showed an asymmetric elliptical distribution on the constellation plane as we have predicted in our derivations (Supplement, Note 3). Each received constellation point represents the recovered $\alpha_s'$ in Eq.(7) normalized by $\sqrt{\langle n_s\rangle/2}$. With $\langle n_s\rangle$ increasing, the fluctuations surrounding each constellation point decreases. Moreover, the simulation results (Supplement, Note 4) for the constellations with $CSPR = 10$ dB, $\langle n_s\rangle$ varying from 20 to 200 are concluded in

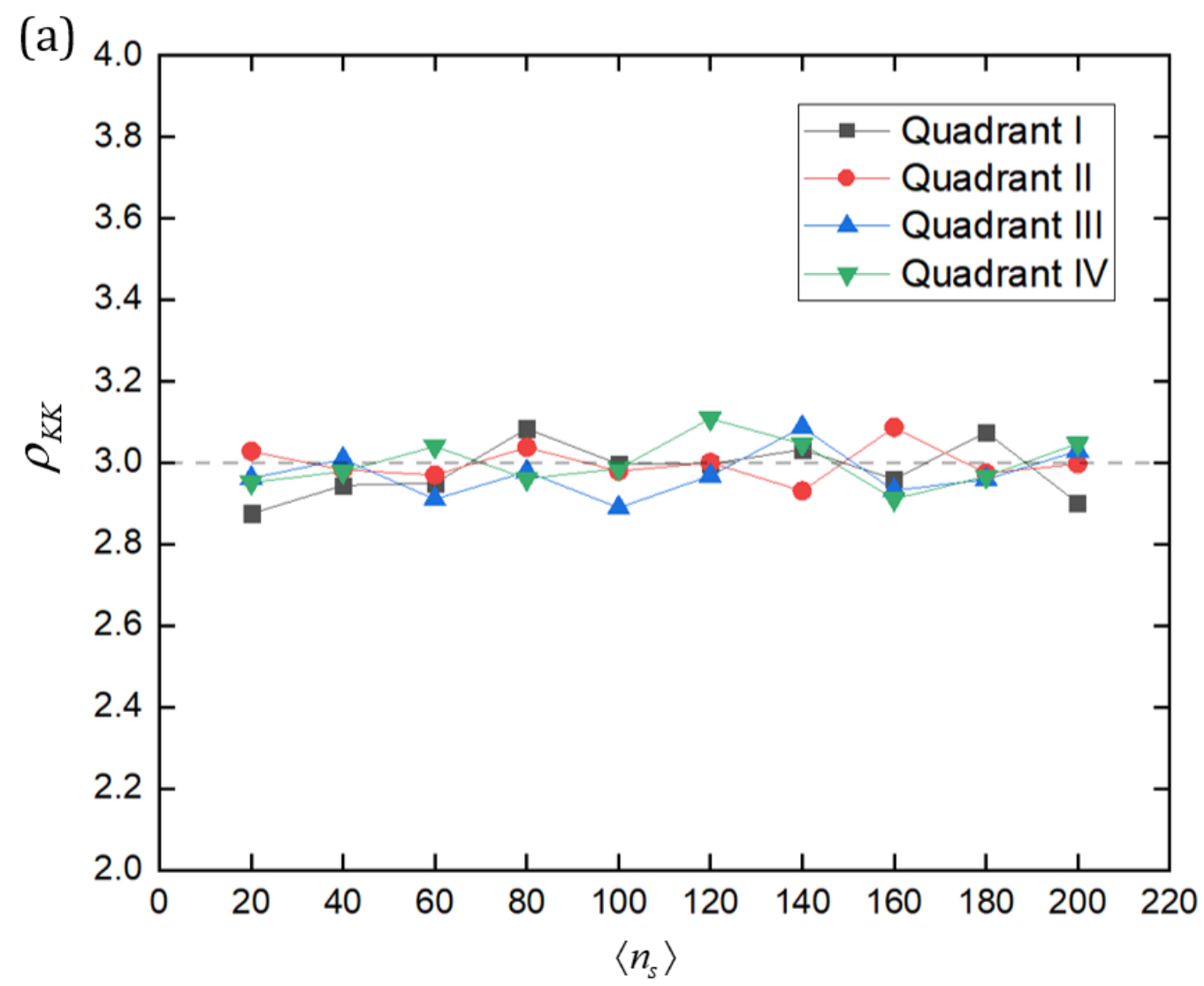

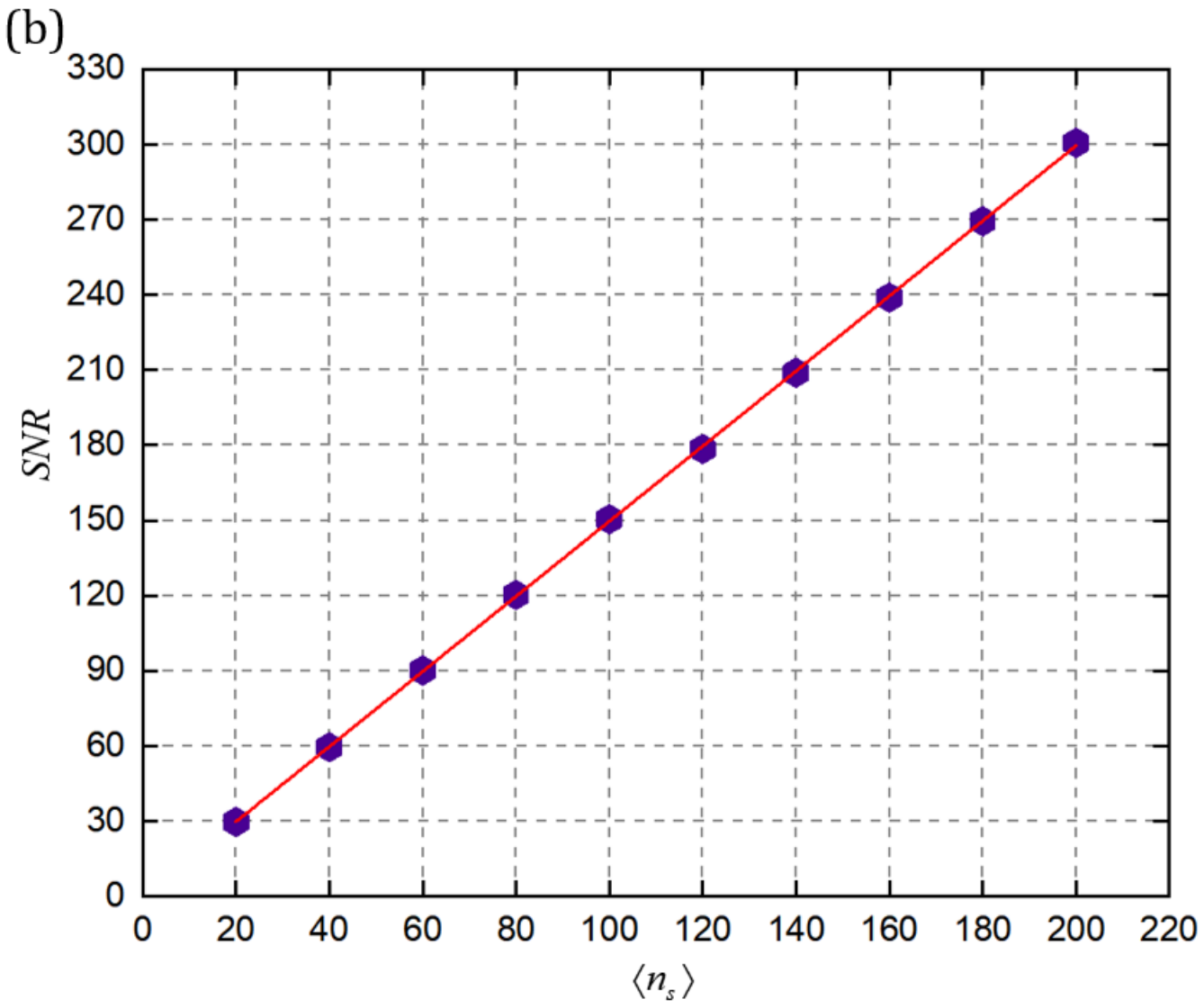

**Fig. 5.** The analysis results for the constellations with $CSPR = 10\text{dB}$ & $\langle n_s\rangle$ varying from 20 to 200. (a) $\rho_{KK}$ of each quadrant;(b)The relation between $SNR$ and $\langle n_s\rangle$.

Fig. 5 (a) and Fig. 5 (b), in which the square of the ratio of the major axis to the minor one of fluctuation ellipse is defined as $\rho_{KK}$.

Fig. 5 (a) shows $\rho_{KK}$ in four quadrants calculated by principal component analysis method provided by MATLAB. Fig. 5 (b) shows the calculated $SNR$ of the received QPSK signals based on KK receiver for different $\langle n_s\rangle$. As shown in Fig. 5 (a), the $\rho_{KK}$ of the simulation results is almost consistent with the predicted value $3/1$. While Fig. 5 (b) indicates that for KK receiver, the $SNR$ of the signal is proportional to its photon number $\langle n_s\rangle$, which shows approximately a linear correlation of $SNR = \frac{3}{2}\langle n_s\rangle$. The simulative results of Fig. 5 (a) and Fig. 5 (b) are consistent with our analytic conclusions Eq. (13), Eq. (14) and Eq. (16).

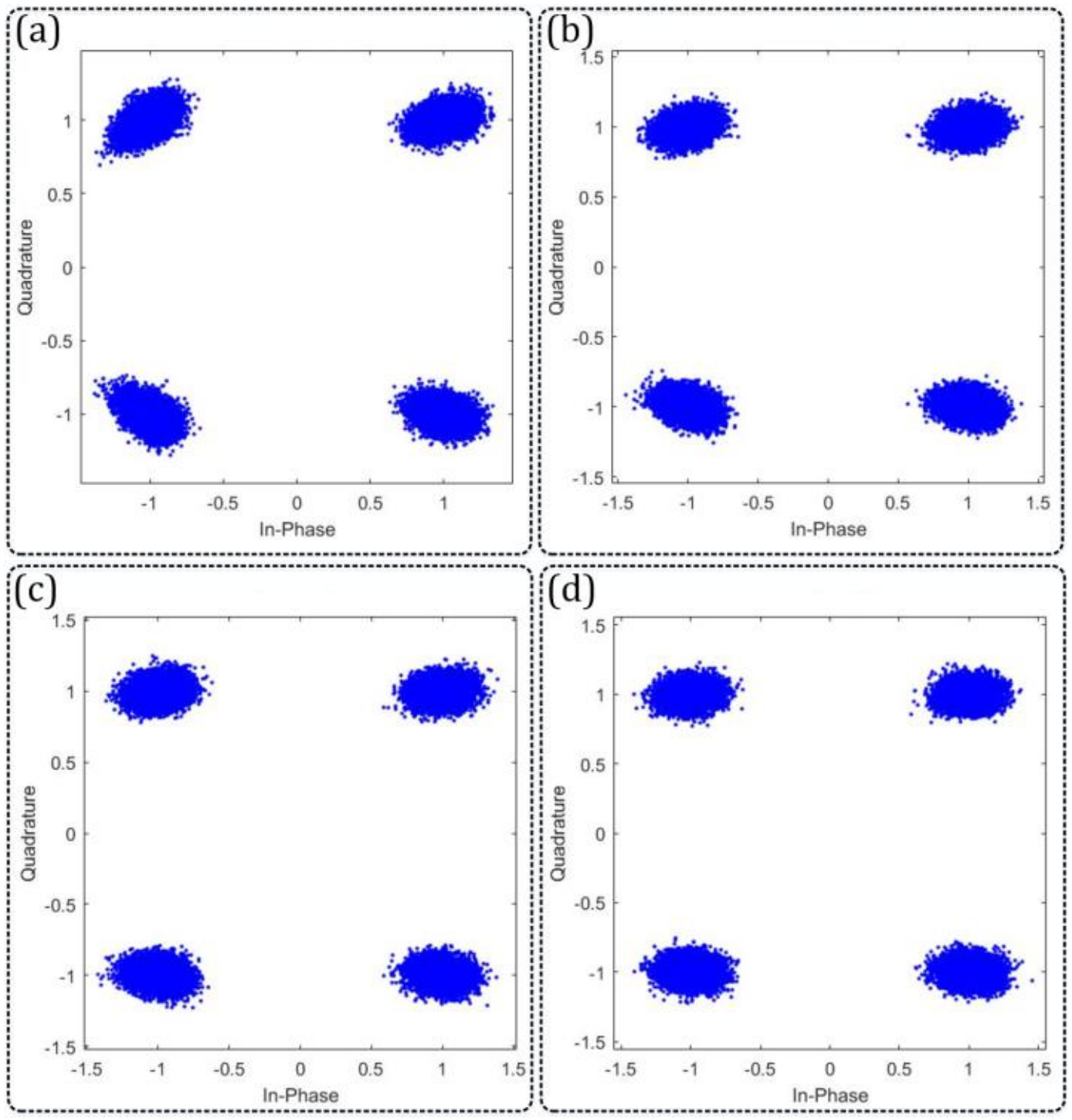

**Fig. 6.** The constellations of $\alpha_s'$ for different $CSPR$ values with $\langle n_s \rangle = 100$ and $\omega_{IF} t \text{ mod } T_{IF} = 0$ . (a) $CSPR = 5$ dB ; (b) $CSPR = 10$ dB ; (c) $CSPR = 15$ dB ; (d) $CSPR = 20$ dB .

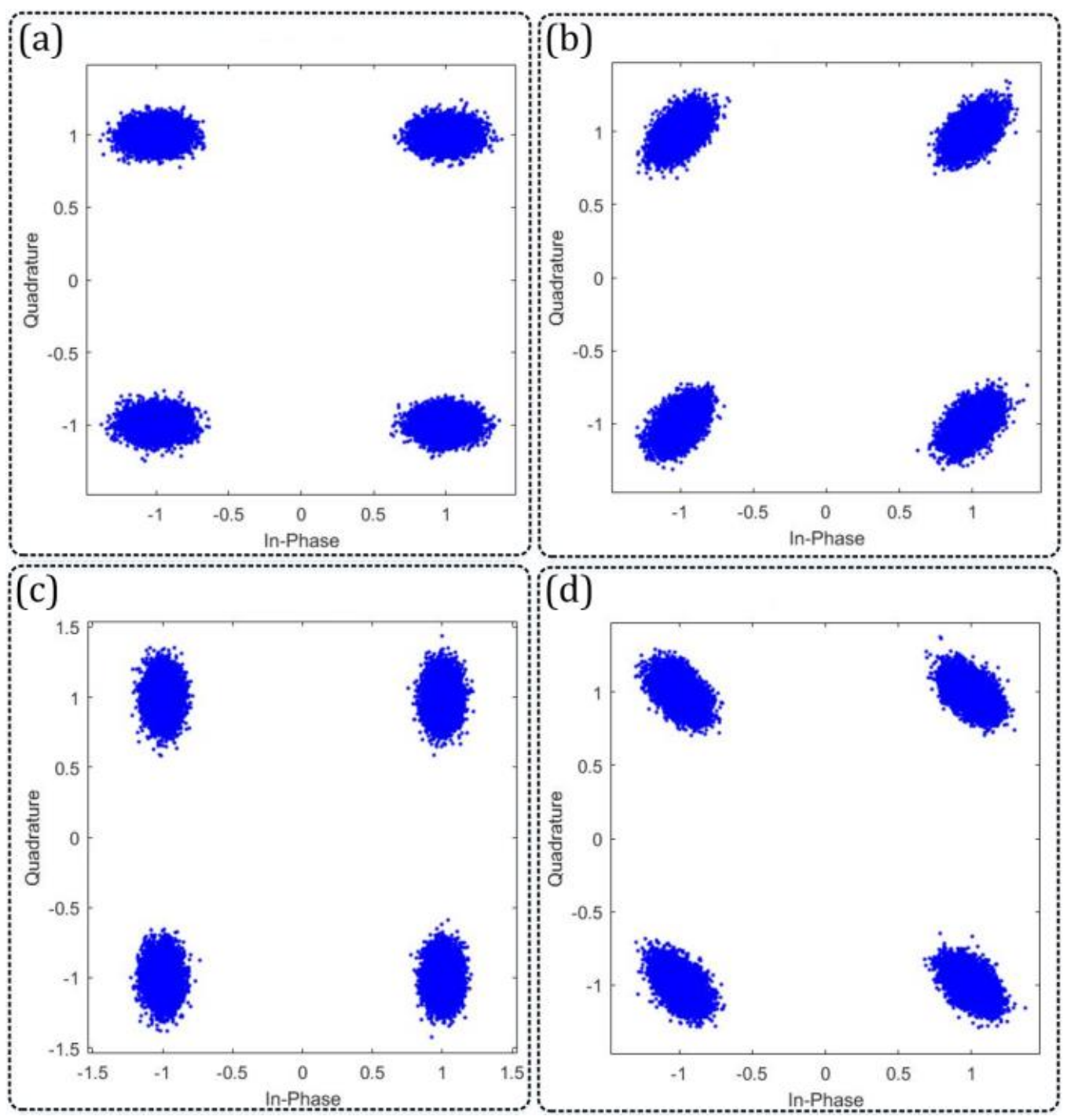

**Fig. 7.** The received constellations of different decision time $t$ with $\langle n_s \rangle = 100$ and $CSPR = 30\text{dB}$ . (a) $\omega_{IF} t \text{ mod } T_{IF} = 0$ ; (b) $\omega_{IF} t \text{ mod } T_{IF} = \frac{\pi}{4}$ ; (c) $\omega_{IF} t \text{ mod } T_{IF} = \frac{\pi}{2}$ ; (d) $\omega_{IF} t \text{ mod } T_{IF} = \frac{3\pi}{4}$ .

Fixing $\langle n_s \rangle$ as 100 and $\omega_{IF} t \text{ mod } T_{IF} = 0$ , Fig. 6 illustrates the constellation diagram of the retrieved signal $\alpha_s'$ with the $CSPR$ set as 5 dB, 10 dB, 15 dB and 20 dB, respectively. The horizontal and vertical coordinates of the constellations respectively represent the in-phase and quadrature components of $\alpha_s'$ recovered through Eq.(7).

As shown in Fig. 6, with the $CSPR$ increasing, the direction of the major axis of the fluctuation ellipse in each quadrant, tends to be consistent with the direction of the carrier as predicted at the end of Section IV. This implies that when the $CSPR$ is large enough ( $(1-\varepsilon)\langle n_L \rangle \gg \varepsilon \langle n_s \rangle$ holds), the distribution of the fluctuations will be determined by $\omega_{IF} t$ almost completely.

For the simulation of the four cases predicted in Fig. 4, the $CSPR$ is set as 30 dB to meet the condition $(1-\varepsilon)\langle n_L \rangle \gg \varepsilon \langle n_s \rangle$ . Fig. 7 (a)-(d) show the constellations of $\alpha_s'$ when the decision time $t$ is selected to meet the conditions $\omega_{IF} t \text{ mod } T_{IF} = 0, \frac{\pi}{4}, \frac{\pi}{2} \,\&\, \frac{3\pi}{4}$ , respectively. While $T$ is adjusted accordingly with the varying $t$ to always meet the condition $t = T/2$ . As shown in Fig. 7 (a) and Fig. 7 (c), we can either reduce the in-phase or quadrature fluctuation to only 1/6, while keeping that of the other orthogonal direction at 1/2. In Fig. 7 (b) and Fig. 7 (d), the fluctuation of both in-phase and quadrature components can be reduced to 1/3 at the same time. The results indicate that for high $CSPR$ such as $CSPR \geq 30$ dB , the KK receiver can achieve a high controllability of the fluctuation distribution on the constellation plane. This is implemented by selecting the decision time $t$ according to the accuracy requirements for the specific direction to be measured. It should be noted that our conclusion does not violate Heisenberg uncertainty principle. Only the quantum fluctuation of the component of the selected direction can be reduced to 1/6, while the quantum fluctuation of the corresponding orthogonal component is kept as 1/2. The total quantum fluctuation 2/3 of KK detection is larger than its counterpart 1/2 of homodyne detection.

## 6. CONCLUSION

In summary, we analytically derive the quantum noise of the in-phase and quadrature operators of the retrieved coherent-state signal with the KK receiver. The quantum limit of the $SNR$ of the KK receiver is 3/2 times the expectation value of the signal photon number. Therefore, the KK receiver has a larger $S/N$ limit than that of balanced heterodyne detection. Since the intensity of the MP signal is physically measured, while its phase is calculated from the intensity using Kramers-Kronig relations, KK receiver keeps the tangential noise to 1/3 times that of the radial one, further leading to an asymmetric distribution of the quantum fluctuation of the retrieved signal. The projected variances of the in-phase and quadrature operators are time-varying due to the phase evolution of the intermediate frequency, which provides us a scheme using a high $CSPR$ to measure the component of a specific direction of a coherent state with fluctuations reduced to 1/6, which is only 2/ 3 times that of the selected direction of a coherent state. The analytical conclusions are validated by numerical simulation. This work provides a physical insight of the KK receiver and should enrich the knowledge of electromagnetic noise in quantum optical measurement.

## Funding.

Peng Cheng Laboratory; National Key Research and Development Program of China (2018YFB1801204, 2016YFA0301402); National Natural Science Foundation of China (62271010, U21A20454, 11834010).

## Disclosures.

The authors declare no competing interests.

## Data availability.

Data underlying the results presented in this paper are not publicly available at this time but may be obtained from the authors upon reasonable request.

## Supplemental document.

See Supplement for supporting content.

# Quantum Noise of Kramers-Kronig Receiver : Supplemental Document

FAN ZHANG,[1,4,5] JIAYU ZHENG,[1] HAIJUN KANG,[2] FENGXIAO SUN,[3] QIONGYI HE,[3] AND XIAOLONG SU[2,6]

[1] *State Key Laboratory of Advanced Optical Communication System and Networks, Frontiers Science Center for Nano-optoelectronics, School of Electronics, Peking University, Beijing 100871, China*
[2] *State Key Laboratory of Quantum Optics and Quantum Optics Devices, Institute of Opto-Electronics, Collaborative Innovation Center of Extreme Optics, Shanxi University, Taiyuan 030006, China*
[3] *State Key Laboratory for Mesoscopic Physics, School of Physics, Frontiers Science Center for Nano-optoelectronics & Collaborative Innovation Center of Quantum Matter, Peking University, Beijing 100871, China*
[4] *Peng Cheng Laboratory, Shenzhen 518055, China.*
[5] *fzhang@pku.edu.cn*
[6] *suxl@sxu.edu.cn*

## Supplementary Note 1： The general proof of the relation of Eq. (9) and Eq. (10)

Here we give a general proof of the relation of Eq. (9) and Eq. (10) in the main text. Considering an operator $\hat{I}$ that measures the intensity of the electrical field consisting of $N$ different frequency modes $\hat{A}_u$ ( $u=1,\cdots,N$ ), we have

$$\hat{I}=(\sum_{u=1}^{N}\lambda_u\hat{A}_u)^\dagger(\sum_{u=1}^{N}\lambda_u\hat{A}_u)\ . \tag{S1}$$

$\lambda_u$ is the strength of $\hat{A}_u$ . We have the following commutation relation as

$$\sum_{u=1}^{N}\lambda_u^2=[(\sum_{u=1}^{N}\lambda_u\hat{A}_u),(\sum_{u=1}^{N}\lambda_u\hat{A}_u)^\dagger]=r\ . \tag{S2}$$

Considering the expression that the operator $\hat{I}^2$ minus the normally ordered operator of $\hat{I}^2$ , we have

$$\begin{aligned}
&\hat{I}^2-:\hat{I}^2: \\
&=[(\sum_{u=1}^{N}\lambda_u\hat{A}_u)^\dagger(\sum_{u=1}^{N}\lambda_u\hat{A}_u)]^2 \\
&-[(\sum_{u=1}^{N}\lambda_u\hat{A}_u)^\dagger]^2[(\sum_{u=1}^{N}\lambda_u\hat{A}_u)]^2 \\
&=(\sum_{u=1}^{N}\lambda_u\hat{A}_u)^\dagger[(\sum_{u=1}^{N}\lambda_u\hat{A}_u),(\sum_{u=1}^{N}\lambda_u\hat{A}_u)^\dagger](\sum_{u=1}^{N}\lambda_u\hat{A}_u) \\
&=(\sum_{u=1}^{N}\lambda_u\hat{A}_u)^\dagger(\sum_{u=1}^{N}[\hat{A}_u,\hat{A}_u^\dagger]\lambda_u^2)(\sum_{u=1}^{N}\lambda_u\hat{A}_u) \\
&=r(\sum_{u=1}^{N}\lambda_u\hat{A}_u)^\dagger(\sum_{u=1}^{N}\lambda_u\hat{A}_u)
\end{aligned} \tag{S3}$$

Here $:\hat{I}^2:$ represents the normally ordered $\hat{I}^2$ only including the terms with the creation operators preceding the annihilation operators.

The expectation of $\hat{I}^2-:\hat{I}^2:$ equals to the fluctuations $\langle(\Delta\hat{I})^2\rangle$ in the measurement of $\hat{I}$ . We have

$$\begin{aligned}
\langle(\Delta\hat{I})^2\rangle&=\langle\hat{I}^2-:\hat{I}^2:\rangle \\
&=r\langle(\sum_{u=1}^{N}\lambda_u\hat{A}_u)^\dagger(\sum_{u=1}^{N}\lambda_u\hat{A}_u)\rangle\ . \\
&=r\langle\hat{I}\rangle
\end{aligned} \tag{S4}$$

As for $\hat{h}=\sqrt{1-\varepsilon}(\hat{A}_L+\hat{A}_s'+\hat{A}_i')+\sqrt{\varepsilon}(\hat{A}_s+\hat{A}_i)$ $(\varepsilon\to1)$ in our paper, we have $N=5$ and $r=\sum_{u=1}^{N}\lambda_u^2=3(1-\varepsilon)+2\varepsilon=3-\varepsilon\to2$ .

Therefore, for $\hat{I}(t)_{KK}=k\hat{h}^\dagger\hat{h}$ , we obtain

$$\langle(\Delta\hat{I}(t))^2\rangle_{KK}=2k\langle\hat{I}(t)\rangle_{KK}\ . \tag{S5}$$

Eq. (10) in the main text can thus be readily obtained from Eq. (9) through the relation that is revealed in Eq. (S5).

## Supplementary Note 2：The derivation of Eq. (12)

Here we give the derivation of Eq. (12) in the main text. Note that Eq. (11) can be used to calculate the phase fluctuations.

$$\langle(\Delta\hat{\phi}(t))^2\rangle = \frac{1}{4\pi^2}\langle\left\{\mathcal{P}\int_{-\infty}^{+\infty}\frac{1}{t-t'}\left[\frac{\hat{I}(t')-I(t')}{I(t')}\right]dt'\right\}^2\rangle \tag{S6}$$

Let $t = l\delta t$, $t' = m\delta t$ and $dt' \to \delta t$, then Eq. (S6) can be rewritten as the following discrete form. To bypass the singularity of the integrand, $m \neq l$.

$$\langle(\Delta\hat{\phi}(t))^2\rangle = \lim_{\delta t\to 0}\frac{1}{4\pi^2}\langle\{\sum_{m=-\infty,m\neq l}^{m=+\infty}\frac{1}{l\delta t - m\delta t}\left[\frac{\Delta\hat{I}(m\delta t)}{I(m\delta t)}\right]\}^2(\delta t)^2\rangle \tag{S7}$$

Due to the white noise property, the measurement of $\Delta\hat{I}$ at different instants $m_1\delta t$ and $m_2\delta t$ are irrelevant. Considering the cross terms in the summation in Eq. (S7), we have

$$\begin{aligned}&\lim_{\delta t\to 0}\frac{1}{4\pi^2}\langle\sum_{m_1,m_2}\frac{1}{(l-m_1)(l-m_2)}\frac{\Delta\hat{I}(m_1\delta t)\Delta\hat{I}(m_2\delta t)}{I(m_1\delta t)I(m_2\delta t)}\rangle\\ &=\lim_{\delta t\to 0}\frac{1}{4\pi^2}\sum_{m_1,m_2}\frac{1}{(l-m_1)(l-m_2)}\frac{\langle\Delta\hat{I}(m_1\delta t)\rangle\langle\Delta\hat{I}(m_2\delta t)\rangle}{I(m_1\delta t)I(m_2\delta t)}.\\ &=0\end{aligned} \tag{S8}$$

Therefore, the cross terms vanish and only the square terms remain. We can thus rewrite $\langle(\Delta\hat{\phi}(t))^2\rangle$ as

$$\begin{aligned}\langle(\Delta\hat{\phi}(t))^2\rangle = \lim_{\delta t\to 0}\frac{1}{4\pi^2}&\left[\sum_{m=-\infty}^{m=l-1}\frac{1}{(l-m)^2}\frac{\langle(\Delta\hat{I}(m\delta t))^2\rangle}{I(m\delta t)^2}\right.\\ &\left.+\sum_{m=l+1}^{+\infty}\frac{1}{(l-m)^2}\frac{\langle(\Delta\hat{I}(m\delta t))^2\rangle}{I(m\delta t)^2}\right]\end{aligned}. \tag{S9}$$

We introduce Eq. (9) and Eq. (10) in the main text into Eq. (S9) and consider Eq. (S5) in the derivation of Eq. (12) in the main text. Note that $\sum_{m=1}^{+\infty} 1/m^2 = \pi^2/6$, the final result of Eq. (12) in the main text is thus obtained.

## Supplementary Note 3：The received constellations with $\langle n_s\rangle = 60,100,160,200$ corresponding to Fig. 5

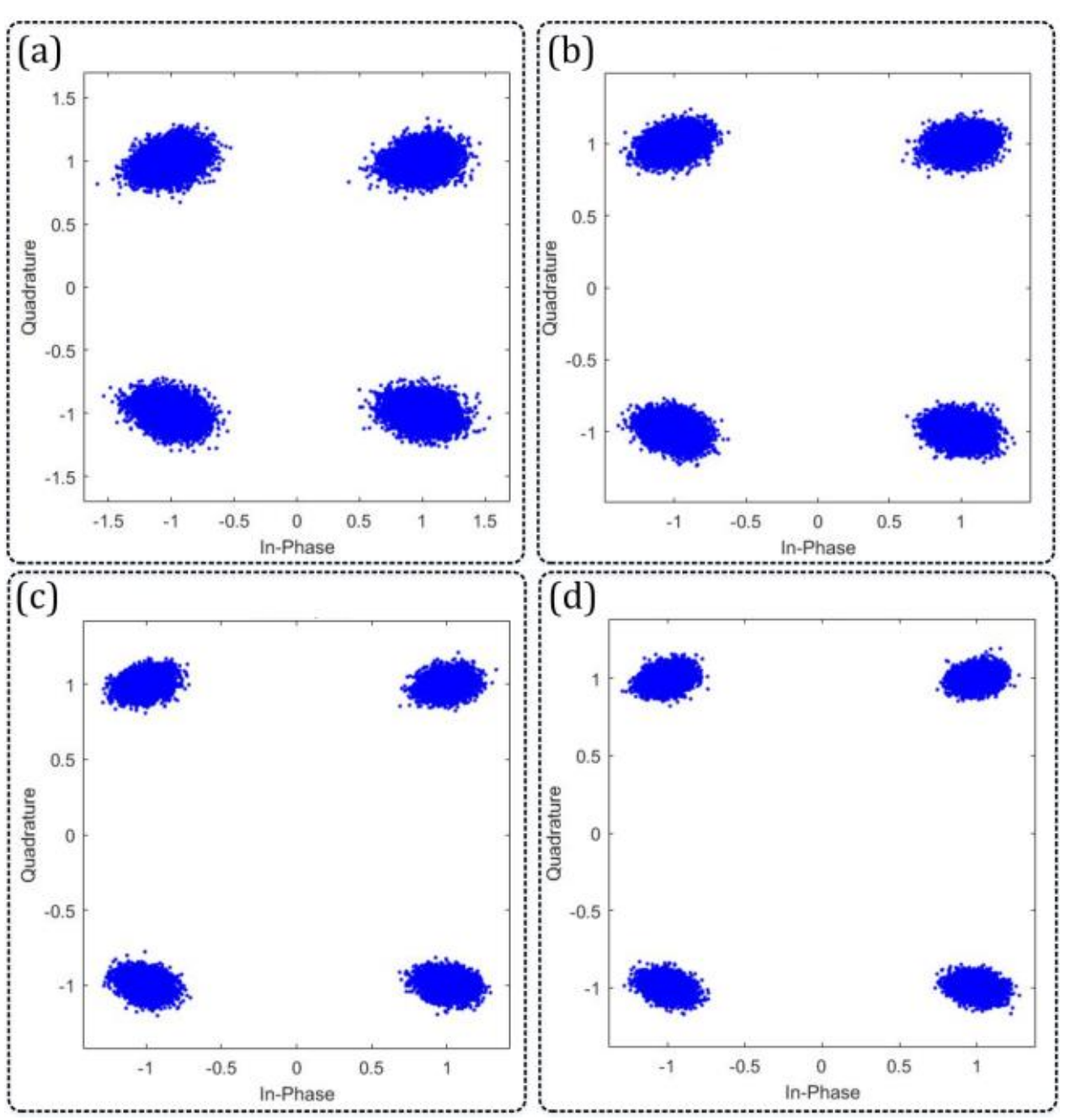

Fig. S1. The received constellations of different $\langle n_s\rangle$ with $CSPR = 10\text{dB}$.

(a) $\langle n_s\rangle = 60$ ; (b) $\langle n_s\rangle = 100$ ; (c) $\langle n_s\rangle = 160$ ; (d) $\langle n_s\rangle = 200$ .

## Supplementary Note 4: $\rho_{KK}$ & $SNR$ varying with $\langle n_s \rangle$ corresponding to Fig. 5

Table. S1. The $\rho_{KK}$ result of each quadrant from 1 to 4 for different $\langle n_s \rangle$

| $\langle n_s \rangle$ | $\rho_{KK}$ of 1-4 quadrant | | | |
|---|---|---|---|---|
| | 1 | 2 | 3 | 4 |
| 20 | 2.87 | 3.03 | 2.96 | 2.95 |
| 40 | 2.94 | 2.98 | 3.01 | 2.98 |
| 60 | 2.95 | 2.97 | 2.91 | 3.04 |
| 80 | 3.08 | 3.04 | 2.98 | 2.96 |
| 100 | 2.99 | 2.98 | 2.89 | 2.99 |
| 120 | 2.99 | 3.0 | 2.97 | 3.11 |
| 140 | 3.03 | 2.93 | 3.09 | 3.05 |
| 160 | 2.96 | 3.09 | 2.93 | 2.91 |
| 180 | 3.07 | 2.97 | 2.96 | 2.97 |
| 200 | 2.90 | 2.99 | 3.03 | 3.05 |

Table. S2. The $SNR$ result for different $\langle n_s \rangle$

| $\langle n_s \rangle$ | 20 | 40 | 60 | 80 | 100 | 120 | 140 | 160 | 180 | 200 |
|---|---|---|---|---|---|---|---|---|---|---|
| $SNR$ | 29.72 | 59.46 | 90.18 | 120.35 | 150.44 | 178.51 | 208.98 | 238.96 | 269.36 | 300.69 |